\documentclass{aastex631}

\usepackage{textcomp}
\usepackage[normalem]{ulem}

\usepackage{amsmath}
\usepackage[utf8]{inputenc}
\usepackage{tabularx}

\newcommand\clearrow{\global\let\rowmac\relax}
\usepackage{booktabs}
\usepackage{hyperref}
\usepackage{multirow}
\usepackage{amsmath}
\usepackage{graphicx} 

\shorttitle{Deep Learning for Generating Photospheric Vector Magnetograms of Solar Active Regions}
\shortauthors{Jiang et al.}

\graphicspath{{./}{figures/}}

\begin{document}

\title{A Deep Learning Approach to Generating Photospheric Vector Magnetograms of Solar Active Regions for SOHO/MDI Using SDO/HMI and BBSO Data}

\author{Haodi Jiang}
\affiliation{Institute for Space Weather Sciences, New Jersey Institute of Technology, University Heights, Newark, NJ 07102, USA;
hj78@njit.edu, wangj@njit.edu, haimin.wang@njit.edu}
\affiliation{Department of Computer Science, New Jersey Institute of Technology, University Heights, Newark, NJ 07102, USA}
\affiliation{Department of Computer Science, Sam Houston State University, Huntsville, TX 77341, USA}

\author{Qin Li}
\affiliation{Institute for Space Weather Sciences, New Jersey Institute of Technology, University Heights, Newark, NJ 07102, USA;
hj78@njit.edu, wangj@njit.edu, haimin.wang@njit.edu}
\affiliation{Center for Solar-Terrestrial Research, New Jersey Institute of Technology, University Heights, Newark, NJ 07102, USA}

\author{Zhihang Hu}
\affiliation{Institute for Space Weather Sciences, New Jersey Institute of Technology, University Heights, Newark, NJ 07102, USA;
hj78@njit.edu, wangj@njit.edu, haimin.wang@njit.edu}

\author{Nian Liu}
\affiliation{Institute for Space Weather Sciences, New Jersey Institute of Technology, University Heights, Newark, NJ 07102, USA;
hj78@njit.edu, wangj@njit.edu, haimin.wang@njit.edu}
\affiliation{Center for Solar-Terrestrial Research, New Jersey Institute of Technology, University Heights, Newark, NJ 07102, USA}

\author{Yasser Abduallah}
\affiliation{Institute for Space Weather Sciences, New Jersey Institute of Technology, University Heights, Newark, NJ 07102, USA;
hj78@njit.edu, wangj@njit.edu, haimin.wang@njit.edu}

\author{Ju Jing}
\affiliation{Institute for Space Weather Sciences, New Jersey Institute of Technology, University Heights, Newark, NJ 07102, USA;
hj78@njit.edu, wangj@njit.edu, haimin.wang@njit.edu}
\affiliation{Center for Solar-Terrestrial Research, New Jersey Institute of Technology, University Heights, Newark, NJ 07102, USA}
\affiliation{Big Bear Solar Observatory, New Jersey Institute of Technology, 40386 North Shore Lane, Big Bear City, CA 92314, USA}

\author{Genwei Zhang}
\affiliation{Institute for Space Weather Sciences, New Jersey Institute of Technology, University Heights, Newark, NJ 07102, USA;
hj78@njit.edu, wangj@njit.edu, haimin.wang@njit.edu}

\author{Yan Xu}
\affiliation{Institute for Space Weather Sciences, New Jersey Institute of Technology, University Heights, Newark, NJ 07102, USA;
hj78@njit.edu, wangj@njit.edu, haimin.wang@njit.edu}
\affiliation{Center for Solar-Terrestrial Research, New Jersey Institute of Technology, University Heights, Newark, NJ 07102, USA}
\affiliation{Big Bear Solar Observatory, New Jersey Institute of Technology, 40386 North Shore Lane, Big Bear City, CA 92314, USA}

\author{Wynne Hsu}
\affiliation{Institute of Data Science,	National University of Singapore, 
Singapore 119077}
\affiliation{Department of Computer Science, School of Computing, National University of Singapore, Singapore 119077}

\author{Jason T. L. Wang}
\affiliation{Institute for Space Weather Sciences, New Jersey Institute of Technology, University Heights, Newark, NJ 07102, USA;
hj78@njit.edu, wangj@njit.edu, haimin.wang@njit.edu}
\affiliation{Department of Computer Science, New Jersey Institute of Technology, University Heights, Newark, NJ 07102, USA}

\author{Haimin Wang}
\affiliation{Institute for Space Weather Sciences, New Jersey Institute of Technology, University Heights, Newark, NJ 07102, USA;
hj78@njit.edu, wangj@njit.edu, haimin.wang@njit.edu}
\affiliation{Center for Solar-Terrestrial Research, New Jersey Institute of Technology, University Heights, Newark, NJ 07102, USA}
\affiliation{Big Bear Solar Observatory, New Jersey Institute of Technology, 40386 North Shore Lane, Big Bear City, CA 92314, USA}

\begin{abstract}
Solar activity is usually caused by the evolution of solar magnetic fields. Magnetic field parameters derived from photospheric vector magnetograms of solar active regions have been used to analyze and forecast  
eruptive events such as solar flares and coronal mass ejections.
Unfortunately, the most recent solar cycle 24 was relatively weak with few large flares, though it is the only solar cycle in which consistent time-sequence vector magnetograms have been available through the Helioseismic and Magnetic Imager (HMI) on board the Solar Dynamics Observatory (SDO) since its launch in 2010.
In this paper, we look into another major instrument, namely the Michelson Doppler Imager (MDI) on board the Solar and Heliospheric Observatory (SOHO) from 1996 to 2010. The data archive of SOHO/MDI covers more active solar cycle 23 with many large flares.
However, SOHO/MDI data only has line-of-sight (LOS) magnetograms. We propose a new deep learning method, named MagNet, to learn from combined LOS magnetograms, $B_{x}$ and $B_{y}$ taken by SDO/HMI along with H$\alpha$ observations collected by the Big Bear Solar Observatory (BBSO), and to generate vector components $B_{x}'$ and $B_{y}'$, which would form vector magnetograms with observed LOS data. 
In this way, we can expand the availability of vector magnetograms to the period from 1996 to present. Experimental results demonstrate the good performance of the proposed method. To our knowledge, this is the first time that deep learning has been used to generate photospheric vector magnetograms of solar active regions for SOHO/MDI using SDO/HMI and H$\alpha$ data.
\end{abstract}


\section{Introduction} 
\label{sec:intro}

Photospheric vector magnetograms of solar active regions (ARs)
 collected by the Helioseismic and Magnetic Imager (HMI) on board the Solar Dynamics Observatory \citep[SDO;][]{2012SoPh..275..207S} play an important role in solar physics.
 They are used by nonlinear force-free field 
 extrapolation methods \citep{Wheatland2000, Wiegelmann2004, Metcalf2005, Schrijver2008a, Sun2012}
 to calculate magnetic energy, which
 provides crucial information concerning 
 an AR's capability of producing eruptive events
including solar flares and coronal mass ejections 
\citep[CMEs;][]{Aschwanden2014a}.
They are also used to derive 
magnetic field parameters
such as those
of Space-weather HMI Active Region Patches \citep[SHARP;][]{SHARP},
which have been used in machine learning-based flare and CME forecasting
\citep{Bobra2015,CMH-2019, 2019ApJ...877..121L,2020ApJ...890...12L,WCT-2020,AW2021}.

Since 2010, consistent time-sequence full disk photospheric vector magnetograms 
have been available through SDO/HMI.
This data set covers a relatively weak solar cycle, namely cycle 24, 
which had fewer large eruptive events,  
and therefore the sampling is not sufficient to understand 
the initiation of solar eruptions
and to make accurate predictions of eruptive events.
Prior to 2010, vector magnetograms were available sporadically for certain ARs from some observatories, such as 
the Big Bear Solar Observatory \citep[BBSO;][]{1999SoPh..184...87D}, 
the Synoptic Optical Long-term Investigations of the Sun \citep[SOLIS;][]{2003SPIE.4853..194K} of the National Solar Observatory (NSO),  
the Imaging Vector Magnetograph \citep[IVM;][]{1996SoPh..168..229M} 
and Haleakala Stokes Polarimeter \citep[HSP;][]{1985SoPh...97..223M}
at the Mees Solar Observatory (MSO),  
the National Astronomical Observatory of Japan \citep[NAOJ;][]{1991LNP...387..320I}, and 
the Solar Optical Telescope Spectro-Polarimeter \citep[SOT-SP;][]{2008SoPh..249..167T} on board the Hinode spacecraft \citep{2007SoPh..243....3K}.
 However, these vector magnetograms were not 
 consistent in the sense that 
 they were not collected on a regular basis.
 On the other hand,
prior to 2010,
full disk
line-of-sight (LOS) magnetograms have been consistently available 
from another source: the Michelson Doppler Imager (MDI) on board the Solar and Heliospheric Observatory \citep[SOHO;][]{1995SoPh..162..129S}. 
This motivates us to develop 
 a deep learning method,
named MagNet, 
to generate consistent time-sequence vector magnetograms 
of all ARs from 1996 to 2010, which covered a more active solar cycle 23.
Our MagNet model is trained by LOS magnetograms, 
$B_{x}$, $B_{y}$ from  SDO/HMI combined with H$\alpha$ observations 
from BBSO.
The validation of our approach uses the overlapping data from
 2010-05-01 to 2011-04-11 when MDI and HMI 
 obtained data simultaneously. There is a good physical reason to use H$\alpha$ 
 as the additional constraint for this research: 
 in the solar atmosphere, magnetic fields and flows are at a frozen-in condition, 
 and therefore the H$\alpha$ fibril structure can provide 
 the direction of magnetic fields in X and Y dimensions \citep{2008SoPh..247..249W, 2021ApJS..256...20J}.

Deep learning has been widely used in 
heliophysics \citep{galvez2019machine,2019ApJ...877..121L, 2020ApJS..250....5J, 2021ApJS..256...20J}
and astronomy \citep{kim2019solar,LWH2021}. 
More recently, deep learning was employed to construct magnetograms and estimate magnetic fields. 
\citet{kim2019solar} generated farside solar magnetograms from STEREO/Extreme UltraViolet Imager
(EUVI) 304-\AA \hspace*{+0.05cm} images using a 
conditional generative adversarial network (cGAN).
The authors trained their cGAN model using pairs of 
SDO/Atmospheric Imaging Assembly (AIA) 304-\AA \hspace*{+0.05cm} images 
and SDO/HMI magnetograms.
\citet{2021A&A...652A.143B} estimated the unsigned radial component of the magnetic field 
from photospheric continuum images
using a convolutional neural network (CNN). \citet{2021ApJ...911..130H, 2022ApJS..259...24H}
performed Stokes inversion to generate vector magnetograms using deep neural networks.
The authors considered both SDO and Hinode missions.
Different from the above approaches, 
our trained MagNet model
takes as input LOS magnetograms from 
SOHO/MDI as well as  
H$\alpha$ images
from BBSO,
and generates as output 
magnetic field components 
$B_{x}'$ and $B_{y}'$.
The generated (predicted) $B_{x}'$ and $B_{y}'$ components along with the LOS components of the magnetic field,
which can be treated as $B_{z}$ components, create vector magnetograms from 1996 to 2010.

The rest of this paper is organized as follows.
Section \ref{sec:observational_data} describes solar observations and data preparation used in this study.
Section \ref{sec:method} presents details of our MagNet model. 
Section \ref{sec:experiment} reports experimental results.
Section \ref{sec:conclusion} presents a discussion and concludes the paper.

\section{Observations and Data Preparation}
\label{sec:observational_data}

MDI \citep{1995SoPh..162..129S}, which is part of the SOHO satellite,
acquires a LOS magnetogram every 96 minutes 
during the period from 1995 to 2011, which covers more active solar cycle 23 with many large flares.
The spatial resolution is 
4\arcsec \hspace*{+0.01in}
and the full disk images are collected on a 1024
$\times$ 1024 detector. 
As a successor of MDI, the HMI instrument, which is part of the SDO mission, provides continuous coverage of full disk Doppler velocities, LOS magnetograms, and continuum proxy images \citep{2012SoPh..275..207S}.
HMI has been operational 
since May 1, 2010,
covering solar cycle 24.
HMI observes the full solar disk at 6173 \AA  \hspace*{+0.015cm} every 12 minutes for a better signal-to-noise ratio. 
The spatial resolution is 
1\arcsec \hspace*{+0.01in} 
and the full disk images are collected on a
4096 $\times$ 4096 detector. 
Photospheric vector magnetograms have been available since the launch of SDO/HMI.
BBSO, which is a ground-based observatory, has provided H$\alpha$ observations
(images) since 1970s \citep{1999SoPh..184...87D}.
BBSO’s full disk H$\alpha$ observations 
are taken every 1 minute, up to 9 hours for one observing day 
at the wavelength of 6563 \AA \hspace*{+0.015cm}.
The spatial resolution is
2\arcsec \hspace*{+0.01in} 
and the full disk images are collected on a
2048 $\times$ 2048 detector.
Unlike satellite-based instruments such as MDI and HMI,
the ground-based full disk telescope at
BBSO sometimes has seeing limitations 
due to unstable conditions of Earth's atmosphere and weather.

In our study, we selected
 full disk H$\alpha$ images based on three criteria.
First, the selected images must be intact. 
Some images are incomplete due to 
(i) operating problems
such as incorrect angles of the telescope, 
loss of focus, 
sudden shaking of the telescope, etc.,
or
(ii) external blocking entities such as birds, airplanes, etc.
These incomplete images were excluded.
Second,
the selected images must show a clear appearance of the solar disk.
Those images with 
unbalanced brightness caused by high and thin clouds were excluded.  
Third, among the images with a very close appearance 
where the identifiable items on the solar disk did not have observable changes, 
the clearest one was selected. 
When items like active regions and filaments are close to the limb, which is darker than the solar center, we 
manually check if the items are clear. 
The images with a very close appearance and less clear items 
were excluded.

In this way we selected full disk H$\alpha$ images 
in the period from 2010-05-01 to 2017-12-31.
For each selected full disk H$\alpha$ image,
we collected its 
temporally closest 
full disk MDI LOS magnetogram,\footnote{The full disk MDI LOS magnetograms were only available from 2010-05-01 to 2011-04-11
in the period between 2010-05-01 and 2017-12-31.} 
full disk HMI LOS magnetogram and 
full disk HMI vector 
magnetograms/components
$B_{x}$, $B_{y}$
where the difference between
the time stamp of the full disk H$\alpha$ image and
the time stamp of a collected 
full disk magnetogram
had to be less than 6 minutes.
The full disk H$\alpha$ image
and the full disk MDI LOS magnetogram
were linearly interpolated to the same spatial size of the full disk HMI LOS magnetogram 
and HMI vector 
magnetograms/components
using the Interactive Data Language (IDL).
These full disk MDI LOS magnetograms 
(HMI LOS magnetograms, HMI vector 
magnetograms/components
$B_{x}$, $B_{y}$ respectively)
were taken directly 
from the mdi.fd\_M\_96m\_lev182
(hmi.M\_720s, hmi.B\_720s respectively) series
at the Joint Science Operations Center (JSOC).\footnote{\url{http://jsoc.stanford.edu/}}

Because we were mainly interested in solar flares in ARs, 
we 
coaligned and cropped
AR patches of 256 $\times$ 256 pixels 
that might produce flares from the 
full disk images
using a two-step coaligning and cropping procedure
written by IDL as follows.\footnote{We
cropped the AR patches (square regions) from the full disk images/magnetograms, and
then all the cropped AR patches
were re-sized to 256 $\times$ 256.
Some of the AR patches 
for the training or testing of the MagNet model
may be from the same active region on different days.}
First, we performed a full disk coalignment via region growing among the 
full disk images.
Then, we cropped and coaligned the AR patches from the 
coaligned full disk images by maximizing
the Pearson correlation coefficient
 \citep[CC;][]{Galton1886,Pearson1895,SS-1990}
 among the AR patches.
The coaligned AR patches with CC below a threshold were discarded.
(In the study presented here, the threshold was
set to 0.9.)
The AR patch size of 256 $\times$ 256 was chosen for 
efficient model training purposes where
the magnetic field strength at the center of each AR patch
was greater than or equal to 1500 Gauss.\footnote{Strictly
speaking, the magnetic field strength of each pixel of the magnetograms collected by HMI and MDI 
is a pixel-area-averaged signal with a unit generally quoted as $Mx/cm^{2}$ \citep{2014SoPh..289.3531C, 2014SoPh..289.3483H}. 
We use Gauss as done in the literature 
\citep{2020ApJ...897L..32R, 2021A&A...652A.143B}
where Gauss is equivalent to $Mx/cm^{2}$ in units.}
A large AR would be segmented into several patches,
all of which were included and used for training.

In preparing the training data,
we selected 2874 full disk H$\alpha$ images 
from 2014-01-01 to 2017-08-04
and their temporally closest 
full disk HMI LOS magnetograms and full disk HMI vector 
magnetograms/components
$B_{x}$, $B_{y}$. 
Each time we coaligned and cropped  
AR patches from four full disk images, 
namely 
a full disk HMI LOS magnetogram, 
a full disk HMI vector magnetogram/component $B_{x}$,
a full disk HMI vector magnetogram/component $B_{y}$ 
and their temporally closest full disk H$\alpha$ image,
using the two-step procedure written by IDL as described above.
As a result of this coaligning and cropping process, we obtained 8442 
coalignments each containing four AR patches.\footnote{Both
the full disk HMI vector magnetograms $B_{x}$, $B_{y}$
and 
the AR patches of $B_{x}$, $B_{y}$
cropped from the full disk magnetograms 
were in the image coordinate system.}
These 8442 coalignments
of AR patches were stored in the training set, denoted Train$\_$HMI,
which was used to train our MagNet model
where all pixels in the AR patches were used to train the model.
The 8442 AR patches of $B_{x}$, $B_{y}$ 
in Train$\_$HMI were used as labels when optimizing the model.

In preparing the test data,
we selected and
constructed two test sets
to evaluate the performance of the trained MagNet model.
The first test set contained 
226 full disk HMI LOS magnetograms
and full disk HMI vector
magnetograms/components 
$B_{x}$, $B_{y}$ from
2017-08-05 to 2017-12-31 and
their temporally closest 
full disk H$\alpha$ images. 
We coaligned and cropped
AR patches of 256 $\times$ 256 pixels 
that might produce flares 
from these full disk images
as done above.
We obtained 
261 coalignments
each containing four AR patches,
which were stored in the Test$\_$HMI set.
The 261 
AR patches of $B_{x}$, $B_{y}$
in Test$\_$HMI were used to
evaluate the AR patches of $B_{x}'$, $B_{y}'$ 
that were predicted (generated) by the trained MagNet model
on Test$\_$HMI.

The second test set contained
115 full disk MDI LOS magnetograms,
full disk HMI LOS magnetograms,
full disk HMI vector magnetograms/components $B_{x}$ and $B_{y}$
from 2010-05-01 to 2011-04-11
and their temporally closest 
full disk H$\alpha$ images. 
During this period in which both MDI's data and HMI's data were available, 
we were able to obtain all
these five types of full disk images.\footnote{Notice
that these 115 collections of full disk images in the second test set were independent and disjoint from the 
226 collections of full disk images in the first test set due to their
totally different collecting periods.
Specifically, the first test set covered the period from 
2017-08-05 to 2017-12-31 whereas the second test set covered the period from 2010-05-01 to 2011-04-11.
The training set, Train$\_$HMI, covered the period from
2014-01-01 to 2017-08-04.}
Each time we 
coaligned and cropped
 AR patches from five full disk images, namely 
 a full disk MDI LOS magnetogram,
a full disk HMI LOS magnetogram,
a full disk HMI vector magnetogram/component $B_{x}$,
a full disk HMI vector magnetogram/component $B_{y}$ and
their temporally closest full disk H$\alpha$ image,
using the two-step procedure
written by IDL as described above.
We obtained 
26 coalignments
each containing five AR patches, 
which were stored in the Test$\_$MDI set.  
The 26 AR patches of $B_{x}$, $B_{y}$
in Test$\_$MDI
were used to 
evaluate 
the AR patches of $B_{x}'$, $B_{y}'$
that were predicted (generated) by the trained MagNet model
on Test$\_$MDI.
Please note that, our training set (Train$\_$HMI) and test sets
(Test$\_$HMI and Test$\_$MDI)
covered different time periods, and therefore were disjoint.
Thus, the trained MagNet model can make 
predictions on data that it has never seen before.

\section{Methodology} 
\label{sec:method}

\subsection{The Workflow of MagNet}

\begin{figure}
		\centering
		\includegraphics[width=6.8in]{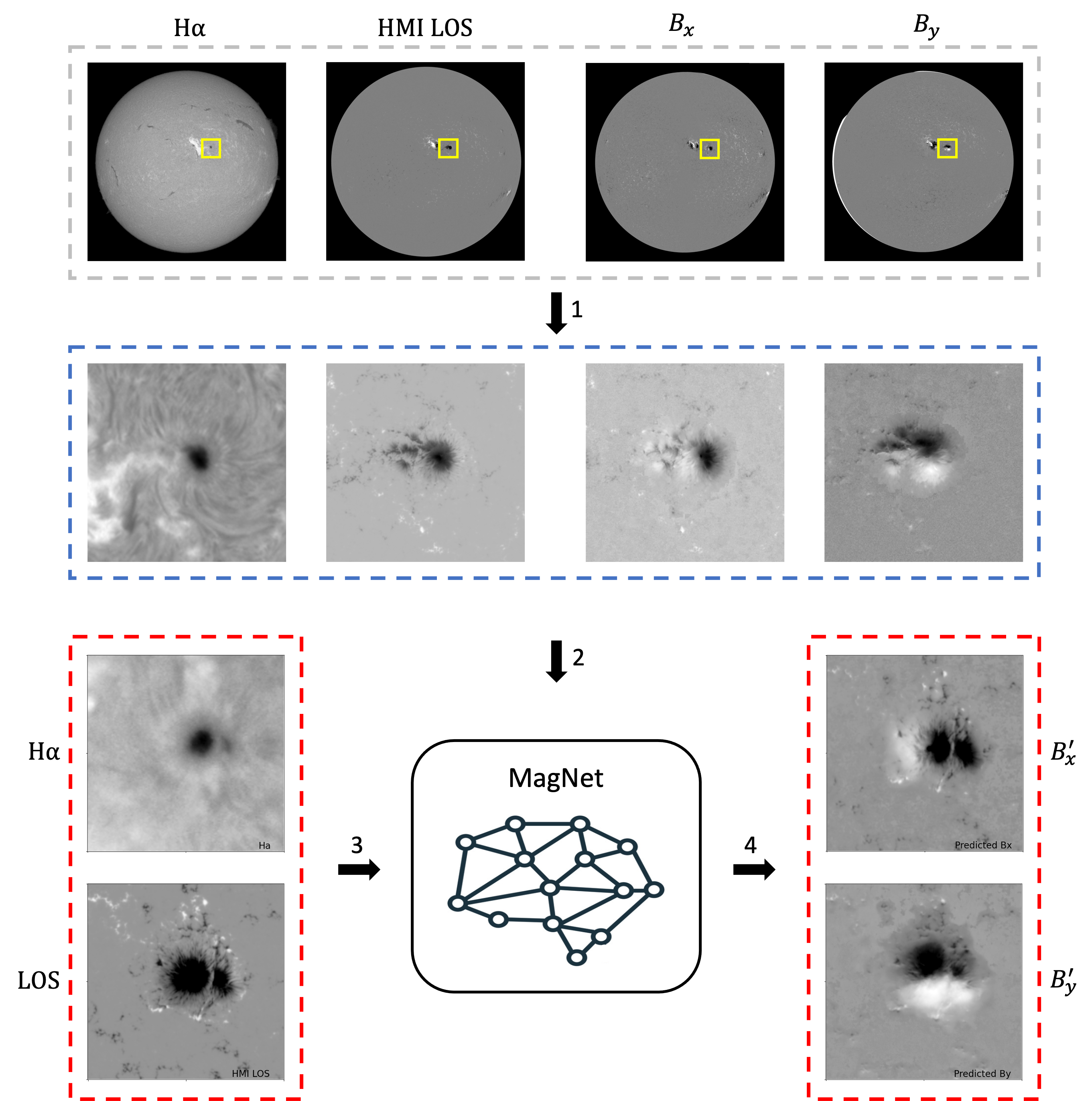}
		\caption{The workflow of MagNet. 
        During training, we coalign and crop AR patches of 256 $\times$ 256 pixels
        from four full disk images (step 1) and use the coaligned AR patches 
        to train the MagNet model (step 2).
        During testing/prediction,
        a pair of coaligned AR patches of H$\alpha$ image 
        and HMI/MDI LOS magnetogram 
        is fed to the trained Magnet model (step 3),
        which generates the pair of AR patches of $B_{x}'$ and $B_{y}'$ that correspond to the input test data
        (step 4).}
		\label{fig: workflow}
\end{figure}

Figure \ref{fig: workflow} illustrates the workflow of MagNet. 
Collected full disk images including 
BBSO H$\alpha$ images, HMI LOS magnetograms, 
and HMI vector components $B_{x}$ and $B_{y}$ 
are shown in the gray dashed box
where an AR patch
in a full disk image is 
highlighted in a small yellow box.
We process these full disk images by 
coaligning and cropping the 
AR patch
out of each full disk image to
produce coaligned
AR patches of 256 $\times$ 256 pixels 
enclosed in the blue dashed box (step 1).
During training, pairs of 
coaligned AR patches of 
BBSO H$\alpha$ images 
and HMI LOS magnetograms 
are fed to the MagNet model where 
the corresponding 
AR patches of 
HMI vector components $B_{x}$ and $B_{y}$,
treated as true $B_{x}$ and $B_{y}$,
are used as labels (step 2).
During testing/prediction, the test data containing a pair of 
coaligned 
AR patches of 
BBSO H$\alpha$ image and HMI/MDI LOS magnetogram,
shown in the left red dashed box,
is fed to the trained MagNet model (step 3).
The model predicts (generates) the pair of 
AR patches of 
$B_{x}'$ and $B_{y}'$,
shown in the right red dashed box, 
that correspond to the input test data
(step 4).

\subsection{The MagNet Model}

\begin{figure}
		\centering
		\includegraphics[width=7in]{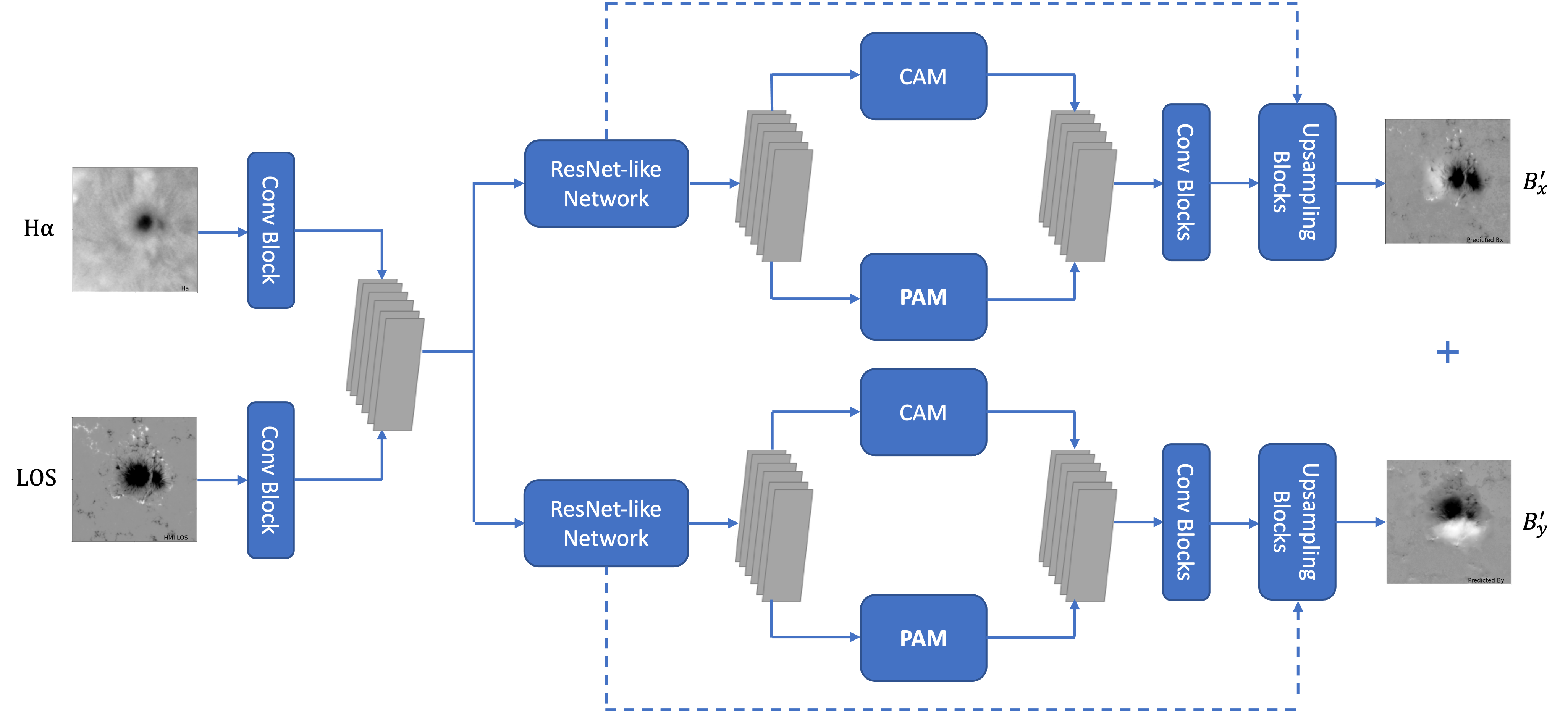}
		\caption{Illustration of the architecture of the MagNet model.
		The model takes as input a pair of 
  coaligned AR patches of
		H$\alpha$ image and HMI/MDI LOS magnetogram, 
		and generates as output 
		the pair of AR patches of
		$B_{x}'$ and $B_{y}'$ that correspond to the input data.
		See text for details of the model architecture.}
		\label{fig: model}
\end{figure}

Generating 
AR patches of 
$B_{x}'$ and $B_{y}'$ 
amounts to solving a regression problem 
because the output of the generating procedure consists of 
real numbers, i.e., magnetic field strengths.
We employ a novel deep learning model containing an   
advanced CNN 
with self-attention \citep{zhao2020exploring} to
solve this regression problem.
Self-attention was
originally developed to learn global dependencies of input data
and was used in machine
translation \citep{NIPS2017_3f5ee243}. 
Later, self-attention was applied to many areas such as 
 image recognition, scene segmentation and image synthesis, etc.
Figure \ref{fig: model} presents the architecture of our deep learning model. 

During training, 
each pair of coaligned 
AR patches of
BBSO H$\alpha$ image
and HMI LOS magnetogram
taken from the training set,
Train\_HMI,
is in turn fed to the MagNet model.
The input images are
first processed by convolutional blocks.
Each convolutional block consists of a convolution layer with batch normalization, followed by a parametric ReLU (PReLU) activation
function \citep{DBLP:conf/iccv/HeZRS15}. 
The feature maps produced by the convolutional blocks are concatenated and sent to two separate paths
where one path generates 
the AR patch of
$B_{x}'$
and the other path generates
the AR patch of
$B_{y}'$.
Each path starts with
a ResNet-like network, 
specifically ResNet-101 \citep{he2016deep} with
a dilated convolution strategy \citep{DBLP:conf/cvpr/0005DSZWTA18} as described in \citet{9154612}.
The output of the ResNet-like network is then
sent to two modules:
the channel attention module (CAM) and 
position attention module (PAM) \citep{fu2019dual}.
CAM and PAM leverage the self-attention mechanisms in the modules 
to better capture 
and transform a wider range of contextual information into local features, 
thus enhancing their representation capability. 
Both CAM and PAM are calculated in a similar way 
where CAM applies the self-attention mechanism to image channels 
while PAM focuses on the calculation of location information.
The outputs of CAM and PAM are combined and 
sent to convolutional blocks, followed by 
upsampling blocks (i.e., 
convolutional blocks with upsampling 
and skip connection
represented as dashed lines in Figure \ref{fig: model} \citep{unet_nature}),
to generate 
the AR patch of 
$B_{x}'$ ($B_{y}'$, respectively).
The AR patches of 
generated $B_{x}'$ and $B_{y}'$ 
are compared with labels
(i.e., the corresponding AR patches of true $B_{x}$ and $B_{y}$).
The weights of neurons in the MagNet model are then updated to minimize the error (loss) caused by the comparison.
At the end of the training process, the weights are optimized.

During testing/prediction, the trained MagNet model
takes as input a pair
of coaligned
AR patches of 
BBSO H$\alpha$ image
and HMI LOS magnetogram 
(MDI LOS magnetogram, respectively) 
taken from the
Test\_HMI (Test\_MDI, respectively) set
and generates as output 
the pair of AR patches of
$B_{x}'$ and $B_{y}'$ that correspond to the input test data.
The AR patches of generated $B_{x}'$ and $B_{y}'$
are evaluated by
the corresponding AR patches of 
 HMI vector components $B_{x}$ and $B_{y}$,
treated as true $B_{x}$ and $B_{y}$,
in Test\_HMI (Test\_MDI, respectively).

For many of the AR patches used in our study, a large portion of each of them has small magnetic field strengths
($\leq$ 200 Gauss).
Relatively few pixels in an AR patch
have large magnetic field strengths ($>$ 200 Gauss).
To tackle this imbalanced problem in our datasets,
we employ a novel weighted $L_{w}^{p}$ loss of a pixel $p$, defined as:
\begin{equation}
L_{w}^{p}(s', s) = |\frac{s}{c}||s' - s|.
\label{lw_p}
\end{equation}
Here, $s'$ represents the MagNet-generated magnetic field strength at $p$,
$s$ represents the true magnetic field strength at $p$, and
$c$ is a threshold.
(In the study presented here, the threshold was set to 0.95.) 
The absolute difference between $s'$ and $s$ at $p$, usually reflected by the $L_{1}$ loss,
is multiplied by a weight, $|\frac{s}{c}|$.
This suggests that a pixel $p$ with a larger (smaller, respectively) 
magnetic field strength 
yield a larger (smaller, respectively) $L_{w}^{p}$ loss.

The weighted loss between  
an AR patch, $A$, of 
generated $B_{x}'$
and 
the corresponding AR patch of 
 true $B_{x}$,
denoted $L_{w}(B_{x}', B_{x})$, is defined as:
\begin{equation}
L_{w}(B_{x}', B_{x}) = \frac{1}{N} 
\sum_{p \in A} 
L_{w}^{p}(s', s),
\label{Lw_loss}
\end{equation} 
where $N = 256 \times 256 = 65536$ is the total number of pixels 
in $A$. 
The weighted loss between 
an AR patch of
generated $B_{y}'$
and 
the corresponding AR patch of
true $B_{y}$,
denoted $L_{w}(B_{y}', B_{y})$,
is defined similarly.

The total loss, denoted $L_{MagNet}$, 
is then defined as
the sum of $L_{w}(B_{x}', B_{x})$ and $L_{w}(B_{y}', B_{y})$, 
as shown in Equation (\ref{total_loss}) below:
\begin{equation}
L_{MagNet} = L_{w}(B_{x}', B_{x}) + L_{w}(B_{y}', B_{y}). 
\label{total_loss}
\end{equation}

The training of MagNet was done by applying 
the adaptive moment estimation (Adam) optimizer \citep{balles2018dissecting,zou2019sufficient}
to minimize $L_{MagNet}$ with 100 epochs on an NVIDIA A100 
GPU.\footnote{Our
MagNet model is coded by Python and TensorFlow. 
The code and pretrained model are 
available at \url{https://nature.njit.edu/solardb/magnet}. 
The full disk H$\alpha$ images 
can be downloaded from 
\url{http://www.bbso.njit.edu/Research/FDHA/}.  
There are two types of images, *fl* and *fr*, from the above BBSO site.
In this study,
we use the *fl* images.}

\section{Experiments and Results} 
\label{sec:experiment}

\subsection{Evaluation Metrics}
\label{MagNet_sec: metrics}
We adopt three metrics, namely 
the mean absolute error \citep[MAE;][]{SS-1990}, Pearson correlation coefficient \citep[CC;][]{Galton1886,Pearson1895,SS-1990} 
and \% Within t \citep{2021ApJ...911..130H}
to quantitatively evaluate the performance of MagNet.
The first metric is defined as:
\begin{equation} \label{eq:MAE}
\text{MAE} =\frac{1}{N}\sum_{i=1}^N |s_{i}' - {s}_{i}|,
\end{equation}
where $N = 65536$ is the total number of pixels 
in an AR patch of a vector component, 
and
$s_{i}$  (${s}_{i}'$, respectively) 
denotes the true
(generated, respectively) magnetic field strength
for the $i$th pixel, $1 \leq i \leq 65536$, in the 
vector component.
This metric has a unit of Gauss, 
which is used to quantitatively assess the dissimilarity (distance) between the true magnetic field strengths
and generated magnetic field strengths
in the vector component.

The second metric is defined as:
\begin{equation}
\text{CC} =\frac{\text{E}[(T-\mu_T)(G-\mu_G)]}{\sigma_T \sigma_G},
\end{equation}
where $E(\cdot)$ is the expectation.
$T$ and $G$ denote the true magnetic field strengths 
and generated magnetic field strengths respectively in the vector component.
$\mu_T$ and $\mu_G$ denote the mean of $T$ and $G$ respectively.
$\sigma_T$ and $\sigma_G$ denote the standard deviation of $T$ and $G$ respectively.
CC, which does not have units, has a value between $-1$ and $1$. 
When CC is $-1$ or $1$, there is an exact linear relationship
between $T$ and $G$.
When CC is 0, there is no linear dependency 
between $T$ and $G$.

The third metric 
is the percentage count of well-estimated pixels \citep{2022ApJS..259...24H}, which is defined as:
\begin{equation}
\text{$\%$ Within t}=\frac{M}{N} \times 100\%,
\end{equation}
where $M$ denotes the total number of agreement pixels 
in the 
vector component.
We say the $i$th pixel in the 
vector component
is an agreement pixel if 
$|s_{i}-{s}_{i}'|$ is smaller than a user-specified threshold t.
(In the study presented here, the threshold t was set to
100 Gauss.)
This metric is used to quantitatively assess the similarity between the true magnetic field strengths
and generated magnetic field strengths
in the vector component.

\subsection{Quantitative Evaluation of the MagNet Model on HMI and MDI Data}

Table \ref{tab:MDI_e} presents the three
evaluation metric values of
MagNet based on the data in the 
Test$\_$HMI and Test$\_$MDI sets.
Each triple in the table consists of three values
where the left one is the minimum value,
the middle one is the average value, and
the right one is the maximum value.
The row of Test$\_$HMI($B_{x}'$)
(Test$\_$HMI($B_{y}'$), respectively)
shows the metric values
obtained by generating $B_{x}'$
($B_{y}'$, respectively)
based on the test data in Test$\_$HMI.
The row of Test$\_$MDI($B_{x}'$)
(Test$\_$MDI($B_{y}'$), respectively)
shows the metric values
obtained by generating $B_{x}'$
($B_{y}'$, respectively)
based on the test data in Test$\_$MDI.
The training data were from Train$\_$HMI.

It can be seen from Table \ref{tab:MDI_e} that
the generated $B_{x}'$ ($B_{y}'$, respectively) components
are close to the true 
$B_{x}$ ($B_{y}$, respectively) components
with the average MAE being less than 100 Gauss for both of the HMI and MDI test data.
The average CC is approximately 0.9 for the HMI test data and 
close to 0.8
for the MDI test data.
The average \% Within t is 
approximately 83\% for the HMI test data and 73\% for the MDI test data.
MagNet generally performs better on the HMI test data than on the MDI test data.
This happens due to several reasons.
First, the MagNet model is trained by HMI data, not MDI data.
The spatial resolution of MDI images is 4\arcsec,
which is lower than the resolution, 1\arcsec, of HMI images.
Furthermore, due to the larger cadence (96 minutes) of MDI
compared to the cadence (12 minutes) of HMI,
the time gaps between the MDI images and their coaligned H$\alpha$ images
are often larger than the time gaps
between the HMI images and their coaligned H$\alpha$ images.
As a result, the quality of  
the coalignments of AR patches
for MDI is 
lower than
the quality of 
the coalignments of AR patches
for HMI. 
In what follows we present some sample predictions for HMI and MDI.

\begin{table}
\centering
\caption{Evaluation Metric Values of MagNet Based on the Test Data} 
\label{tab:MDI_e}
\begin{tabular*}{0.9999\textwidth}{@{\extracolsep{\fill}} c c c c c}
\hline
\hline
Test Sets & MAE  & CC  & \% Within t 
\\ \hline
Test$\_$HMI($B_{x}'$) & (38.55, 63.11, 295.92) & (0.7149, 0.9105, 0.9703) & (36.50$\%$, 84.19$\%$, 95.03$\%$) \\
Test$\_$HMI($B_{y}'$) & (41.76, 62.62, 216.13) & (0.7043, 0.8641, 0.9704) & (34.39$\%$, 82.03$\%$, 93.90$\%$) \\
Test$\_$MDI($B_{x}'$) & (51.98, 87.35, 174.85) & (0.6683, 0.7510, 0.8384) & (56.52$\%$, 74.30$\%$, 88.18$\%$) \\
Test$\_$MDI($B_{y}'$) & (48.73, 83.63, 130.58) & (0.6737, 0.7956, 0.8808) & (55.40$\%$, 72.01$\%$, 89.93$\%$) \\ 
\hline
\end{tabular*}
\end{table}

\subsection{Case Studies of the MagNet Model on HMI and MDI Data}

We follow the Mount Wilson (or Hale) classification system for sunspot groups 
\citep{1919ApJ....49..153H,2016ApJ...820L..11J} to 
indicate whether an active region is complex or not.
Specifically, we consider active regions in the $\alpha$ and $\beta$ classes to be simple ARs, and
active regions in the other classes to be complex ARs.

\begin{figure}
	\centering
	\includegraphics[width=6.8in]{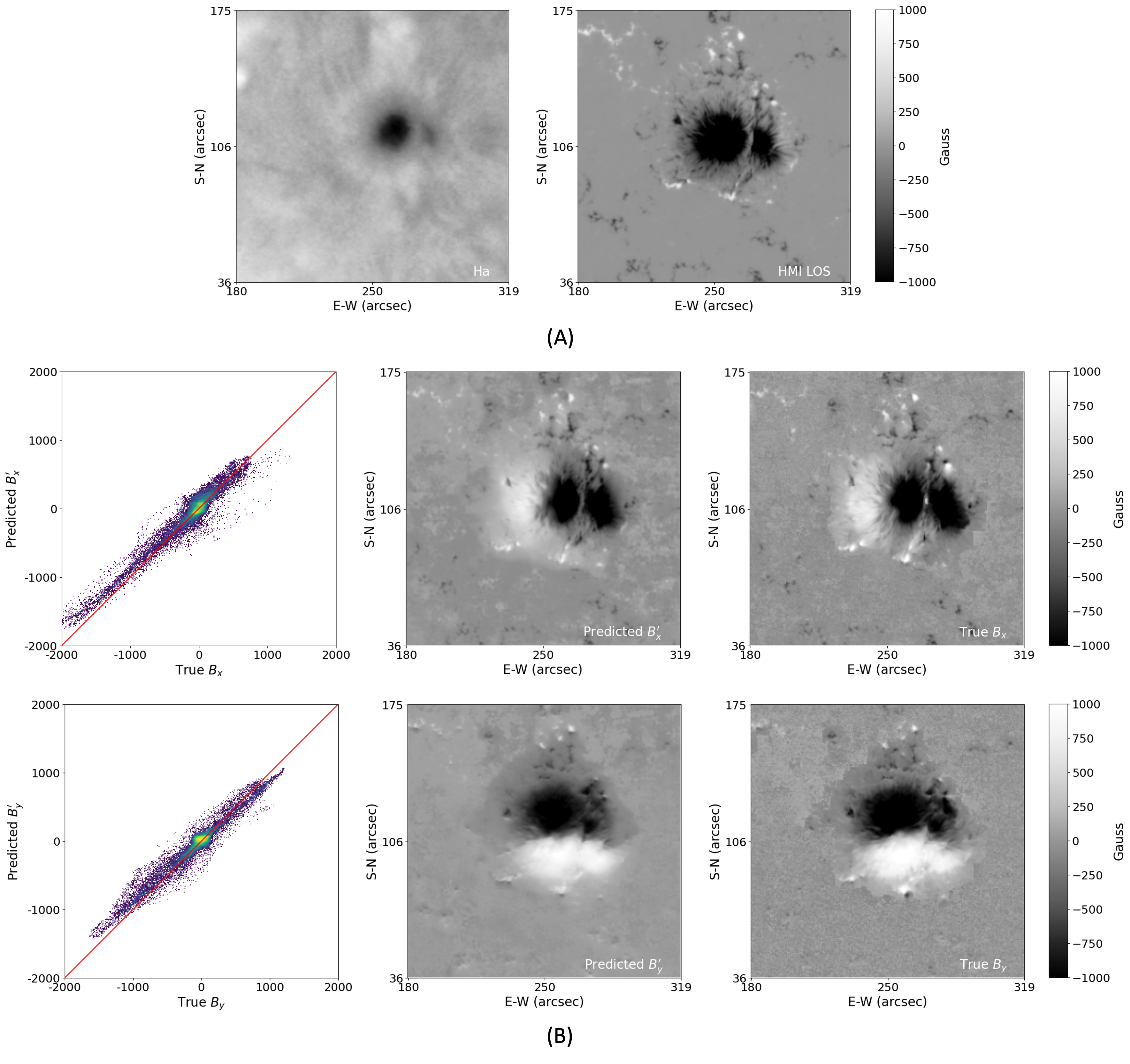}
	\caption{Comparison between MagNet-generated $B_{x}'$, $B_{y}'$
		and true $B_{x}$, $B_{y}$ 
		based on BBSO H$\alpha$ and HMI LOS data
		of AR 12683 on 2017 October 2 21:00:00 UT.
(A) The H$\alpha$ and HMI LOS data inputted to MagNet.
		(B) The output results from MagNet.
  See text for details of the input data and output results.}
	\label{fig:hmi_simple_AR12683}
\end{figure}

\textbf{Generating vector components of AR 12683 based on BBSO H$\alpha$ and HMI LOS data.} 
Figure \ref{fig:hmi_simple_AR12683} presents generated 
$B_{x}'$ and $B_{y}'$ components for AR 12683 
on 2017 October 2 21:00:00 UT
where training data were from Train$\_$HMI.
Figure \ref{fig:hmi_simple_AR12683}(A) shows 
a pair of 
coaligned 256 $\times$ 256 patches of
BBSO H$\alpha$ image and HMI LOS magnetogram
from AR 12683.
This pair of test images is used as input to the trained MagNet model.
Figure \ref{fig:hmi_simple_AR12683}(B)
presents results produced by MagNet with respect to the test data in Figure \ref{fig:hmi_simple_AR12683}(A).
In Figure \ref{fig:hmi_simple_AR12683}(B),
the first column shows
2D histograms which illustrate 
the density of the pixel distribution \citep{2020ApJ...897L..32R},
the second column shows 
the AR patches of
$B_{x}'$ and $B_{y}'$ components generated/predicted by 
MagNet,
and the third column shows
the corresponding AR patches of 
true $B_{x}$ and $B_{y}$ components.
In each 2D histogram, the 
		X-axis represents true $B_{x}$ ($B_{y}$, respectively) and the Y-axis represents generated/predicted $B_{x}'$ ($B_{y}'$, respectively).
		The diagonal line in the 2D histogram
		corresponds to pixels whose generated 
		$B_{x}'$ ($B_{y}'$, respectively) values
		are identical to 
		true $B_{x}$ ($B_{y}$, respectively) values.
AR 12683 was a relatively simple active region.
It can be seen from Figure \ref{fig:hmi_simple_AR12683}(B)
that MagNet works extremely well on this simple AR,
capable of generating vector components that are very close to
true components 
with an MAE of 52.34 Gauss and 57.49 Gauss, 
a CC of 0.9581 and 0.9631, 
and a \% Within t of 87.20$\%$ and 84.72$\%$ 
for $B_{x}$ and $B_{y}$ respectively.

\begin{figure}
	\centering
	\includegraphics[width=6.8in]{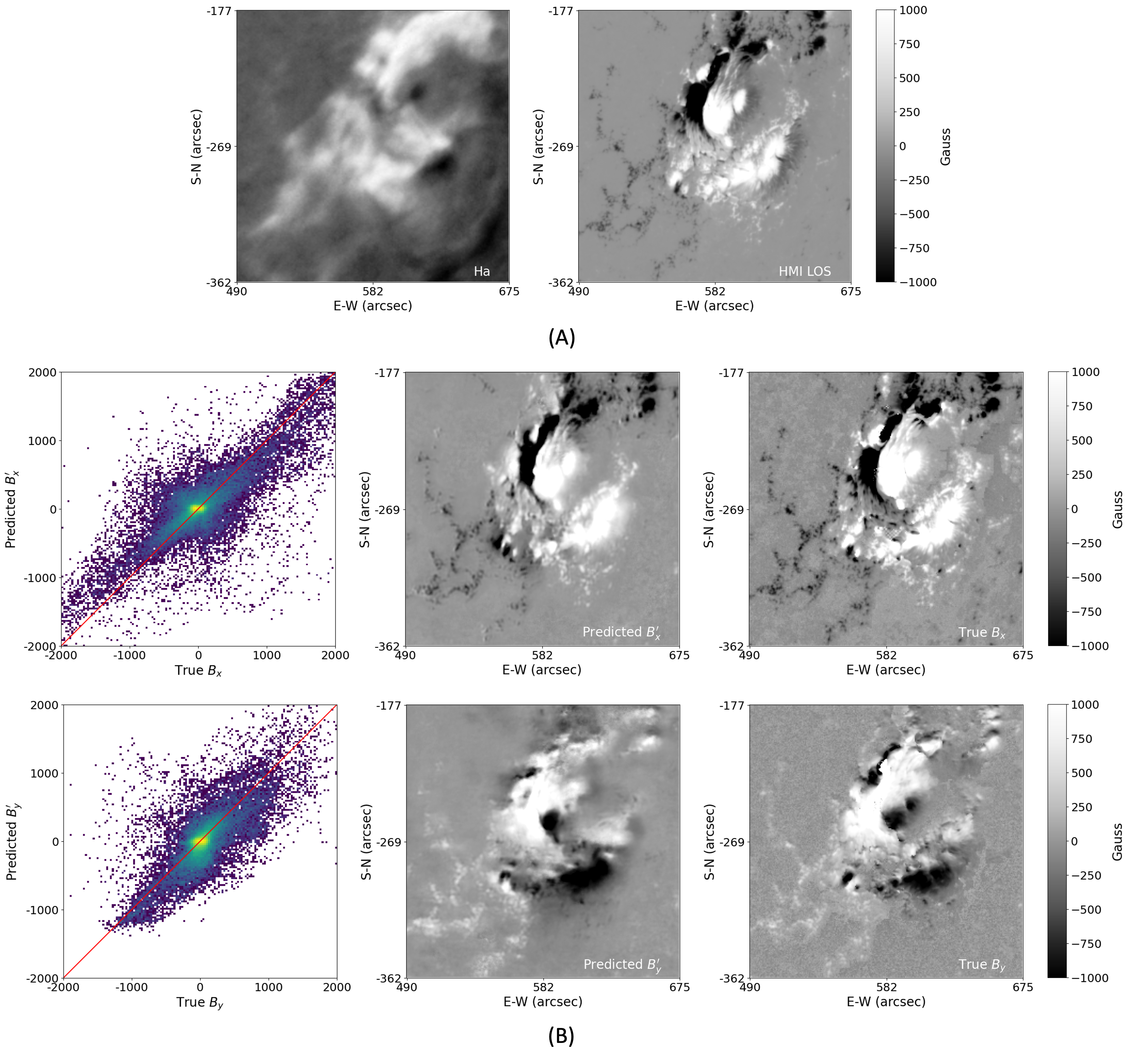}
	\caption{Comparison between MagNet-generated $B_{x}'$, $B_{y}'$
		and true $B_{x}$, $B_{y}$ 
		based on BBSO H$\alpha$ and HMI LOS data
		of AR 12673 on 2017 September 6 19:00:00 UT.
        (A) The H$\alpha$ and HMI LOS data inputted to MagNet.
		(B) The output results from MagNet.
		}
	\label{fig:hmi_complex_AR12673}
\end{figure}
\textbf{Generating vector components of AR 12673 based on BBSO H$\alpha$ and HMI LOS data.}
AR 12673 on 2017 September 6 was a very complex active region.
It was the most flare-productive AR in solar cycle 24,
showing strong magnetic fields in the light bridge and apparent photospheric twist \citep{Wang_2018_inverion}, 
and produced four X-class flares including an X9.3 flare on 2017 September 06.
This active region contained
many pixels with extremely large magnetic field strengths (saturated at the value of 5000 Gauss or larger).
The MagNet model trained by the 8442 
coalignments of AR patches
in the Train$\_$HMI set
described in
Section \ref{sec:observational_data}
did not perform well in AR 12673
on 2017 September 6,
due to the lack of
knowledge of very complex structures and extremely large magnetic field strengths 
such as those in AR 12673 
on 2017 September 6.

It was observed that AR 12673 evolved very rapidly.
There were dramatic changes and significant differences 
during the period between 2017 September 5 and September 7
in which AR 12673 was very complex.
On 2017 September 4, AR 12673 was a very simple active region and
on September 8, it was near the limb.
To enhance the knowledge of MagNet,
we created a new 
training set, denoted Train$\_$HMI$\_$New, 
by manually picking
835 HMI LOS magnetograms from complex ARs in 2015 together with
3 HMI LOS magnetograms and 18 HMI LOS magnetograms from AR 12673 on 2017 September 5 and September 7 respectively.
Due to the rapid daily changes of AR 12673 as explained above,
the information in the new training set, Train$\_$HMI$\_$New, which contained images from 2017 September 5 and September 7,
was disjoint and independent from the information in 
the test images from 2017 September 6.
This new training set contained 
835 + 21 = 856
coalignments each containing four AR patches from 
H$\alpha$ image,
HMI LOS magnetogram,
HMI vector component $B_{x}$ and
HMI vector component $B_{y}$ respectively.

Figure \ref{fig:hmi_complex_AR12673} presents the 
$B_{x}'$ and $B_{y}'$ components for AR 12673 
on 2017 September 6 19:00:00 UT generated by the new MagNet model
trained by Train$\_$HMI$\_$New.
Figure \ref{fig:hmi_complex_AR12673}(A) shows 
a pair of 
coaligned 256 $\times$ 256 patches of
BBSO H$\alpha$ image and HMI LOS magnetogram from AR 12673.
This pair of test images is used as input to the new MagNet model.
Figure \ref{fig:hmi_complex_AR12673}(B) presents
results produced by the new MagNet model with respect to the 
test data in Figure \ref{fig:hmi_complex_AR12673}(A).
It can be seen from Figure \ref{fig:hmi_complex_AR12673}(B)
that the new MagNet model works well on the 
very complex AR 12673 
on 2017 September 6,
capable of generating vector components that are close to
true components 
with an MAE of 145.31 Gauss and 132.33 Gauss, 
a CC of 0.8150 and 
0.7304,
and
a \% Within t of 66.11$\%$ and 63.96$\%$ 
for $B_{x}$ and $B_{y}$ respectively.

\begin{figure}
	\centering
	\includegraphics[width=6.8in]{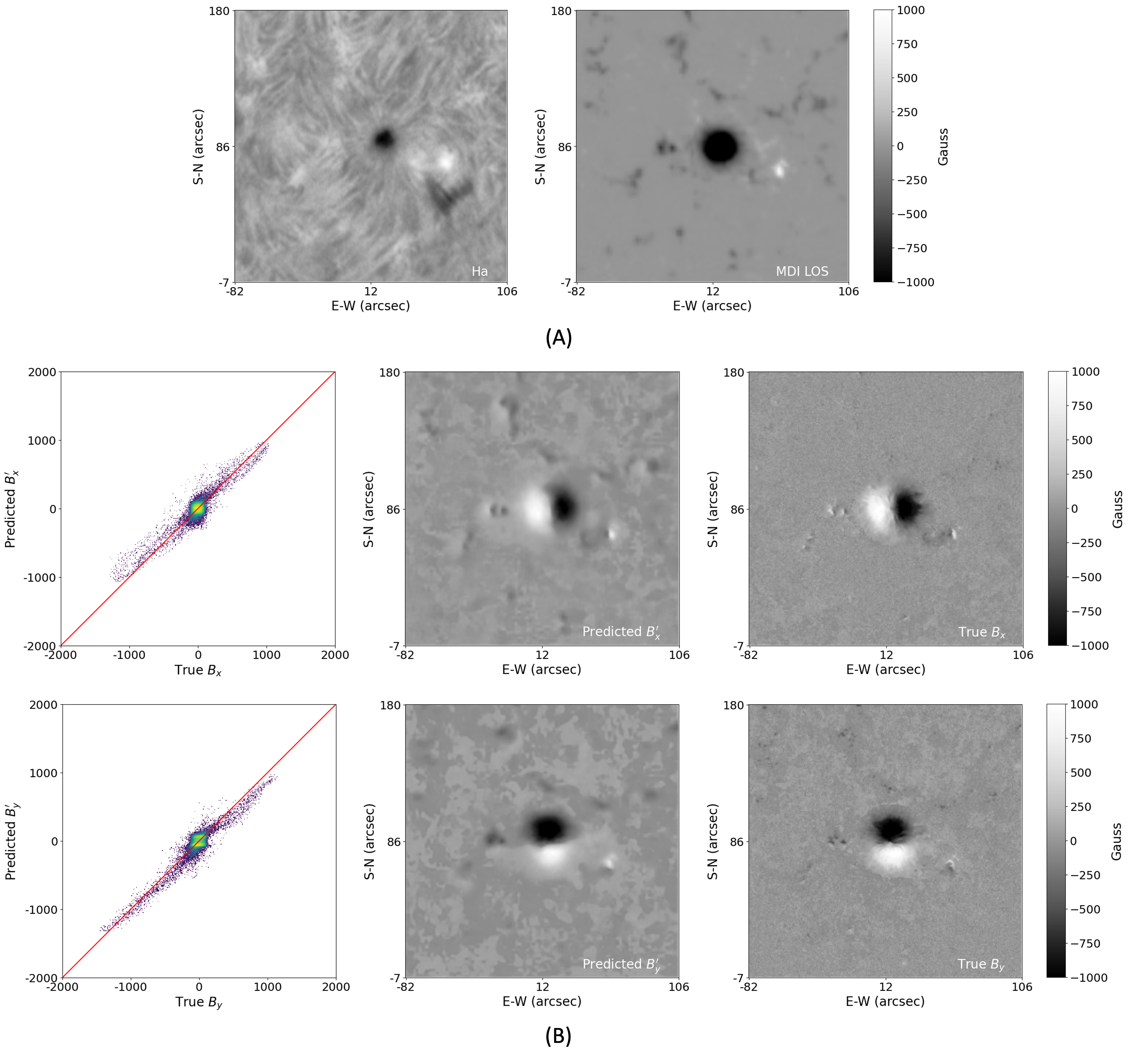}
	\caption{Comparison between MagNet-generated $B_{x}'$, $B_{y}'$
		and true $B_{x}$, $B_{y}$  
		based on BBSO H$\alpha$ and MDI LOS data
		of AR 11101 on 2010 August 30 17:36:00 UT.
        (A) The H$\alpha$ and MDI LOS data inputted to MagNet.
		(B) The output results from MagNet.
		}
	\label{fig:mdi_simple_AR11101}
\end{figure}

\textbf{Generating vector components of AR 11101 based on BBSO H$\alpha$ and MDI LOS data.} 
Figure \ref{fig:mdi_simple_AR11101} presents generated 
$B_{x}'$ and $B_{y}'$ components for AR 11101 
on 2010 August 30 17:36:00 UT
where training data were from Train$\_$HMI.
Figure \ref{fig:mdi_simple_AR11101}(A) shows 
a pair of coaligned 256 $\times$ 256 patches of
BBSO H$\alpha$ image and MDI LOS magnetogram of AR 11101.
This pair of images is used as input to the trained MagNet model.
Figure \ref{fig:mdi_simple_AR11101}(B) presents
results produced by MagNet with respect to the test data in
Figure \ref{fig:mdi_simple_AR11101}(A).
AR 11101 was a relatively simple active region.
It can be seen from Figure \ref{fig:mdi_simple_AR11101}(B)
that MagNet works pretty well on this simple AR,
capable of generating vector components that are similar to
true components
with an MAE of 52.53 Gauss and 
53.16 Gauss, 
a CC of 0.8264 and 
0.8459, 
and a \% Within t of 87.66$\%$ and 
86.33$\%$ 
for $B_{x}$ and $B_{y}$ respectively.
We note both AR 11101 and the AR 12683 described above 
have similar evaluation metric values.
MagNet is trained by HMI data and
the vector components of AR 12683 are generated
based on test data also from HMI
(though the training set and test set are disjoint).
On the other hand, the vector components
of AR 11101
are generated based on test data from MDI.
The fact that AR 11101 and AR 12683  
have similar evaluation metric values
demonstrates the good learning and inference capability of MagNet.

\begin{figure}
	\centering
	\includegraphics[width=6.8in]{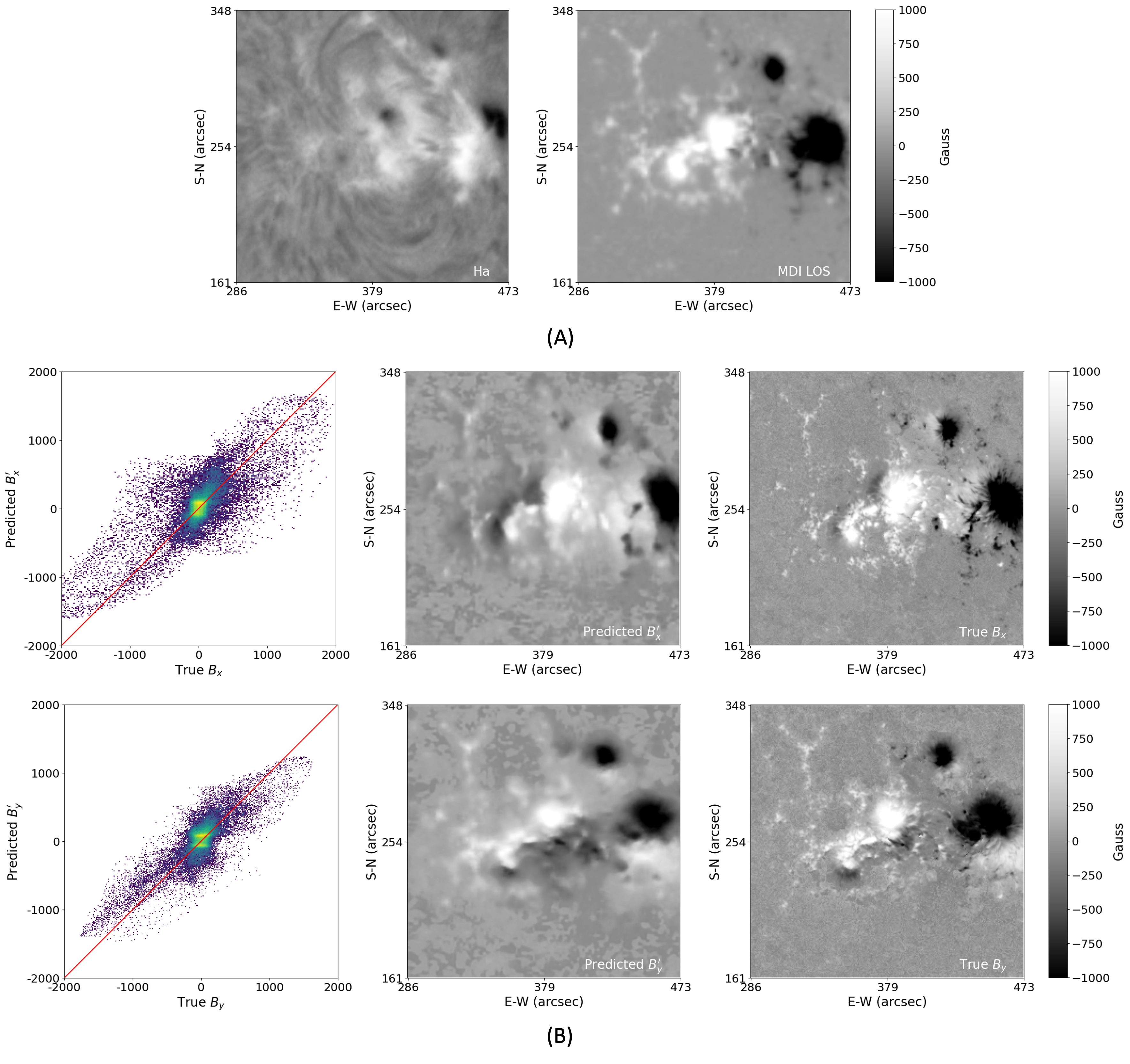}
	\caption{Comparison between MagNet-generated $B_{x}'$, $B_{y}'$
		and true $B_{x}$, $B_{y}$ 
		based on BBSO H$\alpha$ and MDI LOS data
		of AR 11117 on 2010 October 27 22:24:00 UT.
(A) The H$\alpha$ and MDI LOS data inputted to MagNet.
		(B) The output results from MagNet.
		}
	\label{fig:mdi_AR11117}
\end{figure}

\textbf{Generating vector components of AR 11117 based on BBSO H$\alpha$ and MDI LOS data.} 
Figure \ref{fig:mdi_AR11117} presents generated 
$B_{x}'$ and $B_{y}'$ components for AR 11117 
on 2010 October 27 22:24:00 UT
where training data were from Train$\_$HMI.
Figure \ref{fig:mdi_AR11117}(A) shows 
a pair of coaligned 256 $\times$ 256 patches of
BBSO H$\alpha$ image and MDI LOS magnetogram of AR 11117.
This pair of images is used as input to the trained MagNet model.
Figure \ref{fig:mdi_AR11117}(B) presents results produced by MagNet
with respect to the test data in 
Figure \ref{fig:mdi_AR11117}(A).
AR 11117 was a relatively complex active region.
It can be seen from Figure \ref{fig:mdi_AR11117}(B)
that MagNet works reasonably well on this complex AR,
capable of generating vector components that are similar to
true components 
with an MAE of 129.00 Gauss and 114.68 Gauss,
a CC of 0.7514 and 0.8050, 
and a \% Within t of 
61.22$\%$ 
and 58.63$\%$ 
for $B_{x}$ and $B_{y}$ respectively.
Compared to the 
evaluation metric values 
for the relatively simple AR 11101 described above,
the evaluation metric values 
for the more complex AR 11117 are lower.

\section{Discussion and Conclusions} 
\label{sec:conclusion}

We present a new deep learning method (MagNet)
for generating photospheric vector magnetograms of solar active regions for SOHO/MDI using SDO/HMI and BBSO data.
This method allows us to expand the availability of photospheric vector magnetograms to the period from 1996 to present, covering 
solar cycles 23 and 24. 
The vector magnetograms can be used by physics-based methods to calculate magnetic energy and magnetic field parameters useful for 
predicting solar flare activity
\citep{2017ApJ...843..104L,2019ApJ...877..121L,2020ApJ...890...12L}.

Our main results are summarized as follows.
\begin{quote}
1. The experimental results obtained by using BBSO H$\alpha$ observations and HMI magnetograms in the period between 2014-01-01 and 2017-08-04 as training data
demonstrated the good performance of the proposed method. 
Specifically, when using the trained MagNet model to 
generate vector components $B_{x}'$ and $B_{y}'$ based on BBSO H$\alpha$ and HMI LOS data and evaluated by HMI/true vector components $B_{x}$ and $B_{y}$ in the 
Test\_HMI set during the period between
2017-08-05 and 2017-12-31, we obtained 
an average MAE of $\sim$63 Gauss, CC of $\sim$0.9, 
and $\%$ Within t 
of $\sim$83$\%$.
When using the trained MagNet model to generate vector components 
$B_{x}'$ and $B_{y}'$ based on BBSO H$\alpha$ and MDI LOS data and evaluated by the 
HMI/true vector components $B_{x}$ and $B_{y}$
in the Test\_MDI set
from the overlapping period of MDI and HMI between 2010-05-01 and 2011-04-11, we obtained 
an average MAE of $\sim$85 Gauss, 
CC of $\sim$0.8, 
and $\%$ Within t of $\sim$73$\%$.

2. We reported case studies of the MagNet model on HMI and MDI data, which quantitatively and visually demonstrated how our model worked on these data. 
Like other machine learning methods, the performance of MagNet 
depends on training data.
For example, in handling the very complex active region AR 12673,
our original model trained by the Train\_HMI set
does not perform well.
We have to include complex images in the training set 
in order to generate satisfactory vector magnetograms.
Most of the AR patches in the Train\_HMI set are relatively simple; they have small magnetic field strengths with the maximum magnetic field strength being much smaller than 5000 Gauss.
As a consequence, the model trained by 
the Train\_HMI set
performs well in generating vector components with small magnetic field strengths.
It suffers when generating vector components having complex structures with very large magnetic field strengths.
On the other hand,
with the new training data in the Train$\_$HMI$\_$New set, which contains selected complex structures with very large magnetic field strengths, MagNet exhibits good performance as shown in Figure \ref{fig:hmi_complex_AR12673}.

3. When training and testing the MagNet model, we did not impose any threshold on the data.
In other words, all pixels in the original images from the instruments (MDI, HMI, BBSO) were included. 
Imposing thresholds on the data has an impact on the performance of MagNet. 
For example, suppose that, during training,
we only consider pixels whose magnetic field strengths range from $-2000$ Gauss and $+2000$ Gauss \citep{2020ApJ...897L..32R}.
Due to the lack of knowledge of pixels with large magnetic field strengths (e.g., 5000 Gauss) in a complex active region such as
AR 12673, MagNet suffers
when predicting the test images from the complex AR with large magnetic field strengths.

4. MagNet is trained by HMI data and tested on MDI data.
However, HMI and MDI are two different instruments on different observatories with different temporal and spatial resolutions.
One could use deep learning techniques \citep{DBLP:conf/cvpr/MenonDHRR20, DBLP:conf/cvpr/DanierZB22, DBLP:conf/cvpr/RombachBLEO22}
to improve the temporal and spatial resolutions
of the images from the different sources
to enhance the performance of MagNet.
Transfer learning could also be helpful in obtaining knowledge
from one source and applying the knowledge to another source
\citep{DBLP:journals/pieee/ZhuangQDXZZXH21}.

5. Our approach heavily depends on the coalignments 
of AR patches, which are produced by a two-step coaligning and cropping procedure written by IDL.
Coaligning images from the different instruments is a challenge.
Developing more accurate coalignment algorithms
implemented in OpenCV or IDL
may further improve the performance of MagNet.

6. Due to the operation time and condition
limitations,
BBSO did not collect full disk H$\alpha$ images for all the
active regions (ARs) in solar cycle 23.
As a consequence, vector components of those missing ARs would not be generated.
To get full coverage of the ARs in solar cycle 23, one would need to resort to
other sources (e.g., Kanzelhöhe Solar Observatory 
\citep[KSO;][]{1999ASPC..184..314O, 2008CEAB...32....1O}) that provide complementary full disk H$\alpha$ images.
\end{quote}

To our knowledge, MagNet is the first method
capable of generating photospheric vector magnetograms of solar active regions for SOHO/MDI using 
SDO/HMI and H$\alpha$ data. According to our findings and results, the vector components generated by MagNet are reasonably close to ground truths
(observed data).
More research is needed to further improve the quality of the generated data.\  \\

We thank the referee and scientific editor for very helpful
and thoughtful comments.
We also thank the BBSO team for providing the data used in this study. 
The BBSO operation is supported by the New Jersey Institute of Technology
and U.S. NSF grant AGS-1821294. 
This work was supported by U.S. NSF grants AGS-1927578, AGS-1954737, AGS-2149748 and AGS-2228996.
Q.L., J.J., Y.X. and H.W. acknowledge the support of NASA under grants 
80NSSC18K1705,
80NSSC19K0068 and
80NSSC20K1282.

\facilities{Big Bear Solar Observatory, Solar Dynamics Observatory, Solar and Heliospheric Observatory}

\bibliographystyle{aasjournal}

\end{document}